\documentclass[conference]{IEEEtran}
\IEEEoverridecommandlockouts
\usepackage{cite}
\usepackage{amsmath,amssymb,amsfonts}
\usepackage{algorithmic}

\usepackage{algorithm,algorithmic}
\usepackage{graphicx}
\usepackage{optidef}
\usepackage[caption=false]{subfig}
\usepackage{booktabs, makecell, multirow}

\NewExpandableDocumentCommand\mcc{O{1}m}
    {\multicolumn{#1}{c}{#2}}
\usepackage{textcomp}
\usepackage{xcolor}
\makeatletter
\newcommand\fs@betterruled{%
  \def\@fs@cfont{\bfseries}\let\@fs@capt\floatc@ruled
  \def\@fs@pre{\vspace*{5pt}\hrule height.8pt depth0pt \kern2pt}%
  \def\@fs@post{\kern2pt\hrule\relax}%
  \def\@fs@mid{\kern2pt\hrule\kern2pt}%
  \let\@fs@iftopcapt\iftrue}
\floatstyle{betterruled}
\restylefloat{algorithm}
\makeatother
\def\BibTeX{{\rm B\kern-.05em{\sc i\kern-.025em b}\kern-.08em
    T\kern-.1667em\lower.7ex\hbox{E}\kern-.125emX}}
\begin{document}

\title{MAxLM: Multi-Agent Language Model-Based Scheduling and Resource Allocation in MU-MIMO-OFDMA-Enabled Wireless Networks\\
}

\author{\IEEEauthorblockN{Adnan Quadri and Hongxiang Li}
\IEEEauthorblockA{Department of Electrical and Computer Engineering,
University of Louisville, Louisville, KY, USA}
Email: \{adnan.quadri, h.li\}@louisville.edu
}

\maketitle

\begin{abstract}
Wireless networks support multi-user (MU) communication with multiple-input multiple-output (MIMO) and orthogonal frequency-division multiple access (OFDMA) technologies. In the joint MU-MIMO-OFDMA-enabled transmission mode, network throughput can be significantly increased by effectively utilizing the multi-channel resources to schedule numerous wireless users/stations (STAs) simultaneously. In this paper, we study ways to optimize the user scheduling and resource allocation (SRA) for the UL scheduled access (UL-SA) of a joint MU-MIMO-OFDMA-enabled wireless local area network (WLAN). In particular, we propose a multi-agent (MA) framework that utilizes an openly available pretrained small/medium-sized Language Model (xLM) to perform SRA for the UL-SA. To facilitate autonomous SRA using our proposed technique, we introduce the AI-assisted Wireless Systems Engineering and Research (WiSER) platform. We evaluate the performance of MAxLM-optimized SRA for network scenarios with a varying number of STAs and antenna settings on the WLAN Access Point. Numerical results confirm that our proposed technique achieves higher UL-SA throughput than the benchmark techniques.  
\end{abstract}

\begin{IEEEkeywords}
Multi-User Communication, MU-MIMO, OFDMA, Artificial Intelligence, Resource Allocation.
\end{IEEEkeywords}

\section{Introduction}\label{intro}

With the advent of multi-user (MU) communication technologies, such as orthogonal frequency-division multiple access (OFDMA) and multiple-input multiple-output (MIMO), today's Wireless Local Area Networks (WLANs) can serve multiple users concurrently. With OFDMA, sub-channels are grouped into resource units (RUs) to provide multi-channel access over the following RU configurations: nine 26-tone, four 52-tone, two 106-tone, or a single 242-tone RU. With MU-MIMO, concurrent transmissions are supported over the same RU. Therefore, in the joint MU-MIMO-OFDMA transmission mode, the WLAN Access Point (AP) can utilize the multi-channel resources over time, frequency, and space to schedule simultaneous transmissions and increase WLAN throughput. The optimization of wireless user/station (STA) scheduling and resource allocation (SRA) in the joint MU-MIMO-OFDMA-enabled WLAN is a complex combinatorial optimization problem \cite{userGrouping45}. To circumvent complex optimization, Deep Learning \cite{ Raudl}, Deep Reinforcement Learning \cite{userGrouping33}, and Multi-Agent Reinforcement Learning (MARL)-based \cite{userGrouping47, userGrouping46} schemes have been proposed, which are yet to be considered practically amenable.

However, recent advances in generative Artificial Intelligence (AI) pave the way for intent-based network (IBN) management. In \cite{MAESTRO}, a collaborative framework is developed for stakeholders in shared intent-based 6G networks to communicate effectively by leveraging a Large Language Model (LLM)-enabled declaration of intents. In \cite{LiINK}, an LLM is employed to manage 6G-empowered Digital Twin networks.  In \cite{LLMxapp}, a LLM-based resource management scheme is proposed to effectively utilize the physical resource blocks (PRBs) for 5G communication. However, the aforementioned studies focus on either the automation of existing network services or the optimization of wireless network resources using computationally intensive frontier AI-models.  

In contrast, this paper proposes the adoption of a multi-agent framework that leverages an openly available pretrained small/medium-sized Language Model (xLM) to perform SRA for the uplink scheduled access (UL-SA) of a joint MU-MIMO-OFDMA-enabled WLAN. In particular, we address the following challenges of 1) simplifying the complex combinatorial SRA problem, 2) effectively prompting an xLM to perform SRA in a dynamic WLAN environment, and 3) orchestrating the workflow of the xLM-based SRA for the UL-SA. The contributions of the paper are summarized as follows:
\begin{enumerate}
    \item To simplify the optimization of the SRA problem, we adopt a multi-agent framework that utilizes an xLM to perform resource assignments in a decentralized manner. 
    \item To prompt the xLM with updated information on the WLAN environment, we develop the Adaptive Context Management (ACM) procedure that enables the xLM to anticipate the most effective SRA strategy prior to the UL-SA.
    \item To facilitate autonomous SRA, we introduce the AI-assisted Wireless Systems Engineering and Research (WiSER) platform and make it available in \cite{wiser_github}. 
\end{enumerate}

In what follows, Section \ref{nmpf} discusses the joint MU-MIMO-OFDMA-supported WLAN model and the problem formulation. In Section \ref{maxlm} and Section \ref{results}, the proposed technique is discussed and its performance is analyzed, respectively. Finally, Section \ref{conc} concludes the paper. 

\section{Network Model and Problem Formulation}\label{nmpf}

\subsubsection{Network Model} 
\begin{figure}[!t]
\captionsetup[subfloat]{labelformat=simple}
\centering
\subfloat[]{\includegraphics[width=0.5\columnwidth]{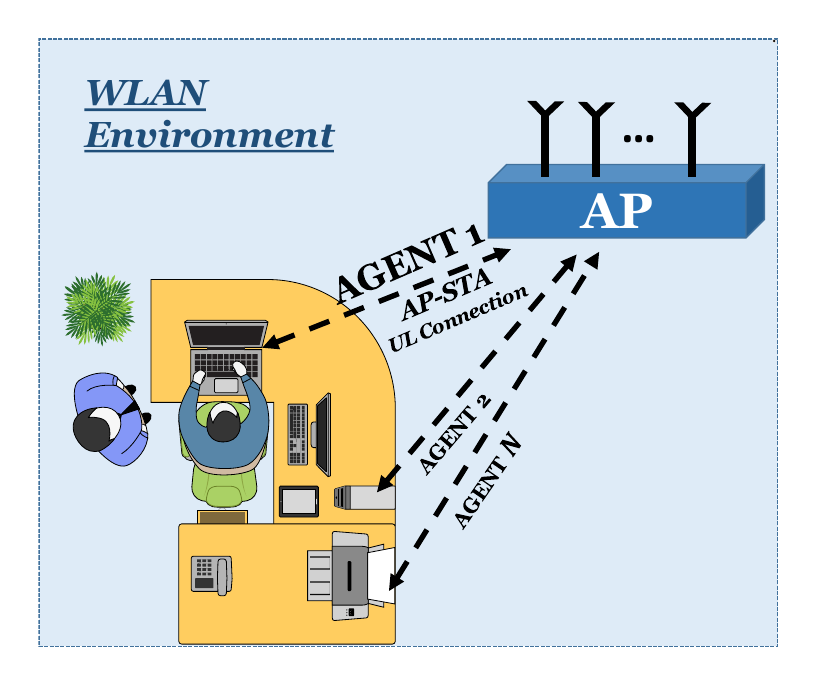}%
\label{wlan_env}}
\hfil
\subfloat[]{\includegraphics[width=0.5\columnwidth]{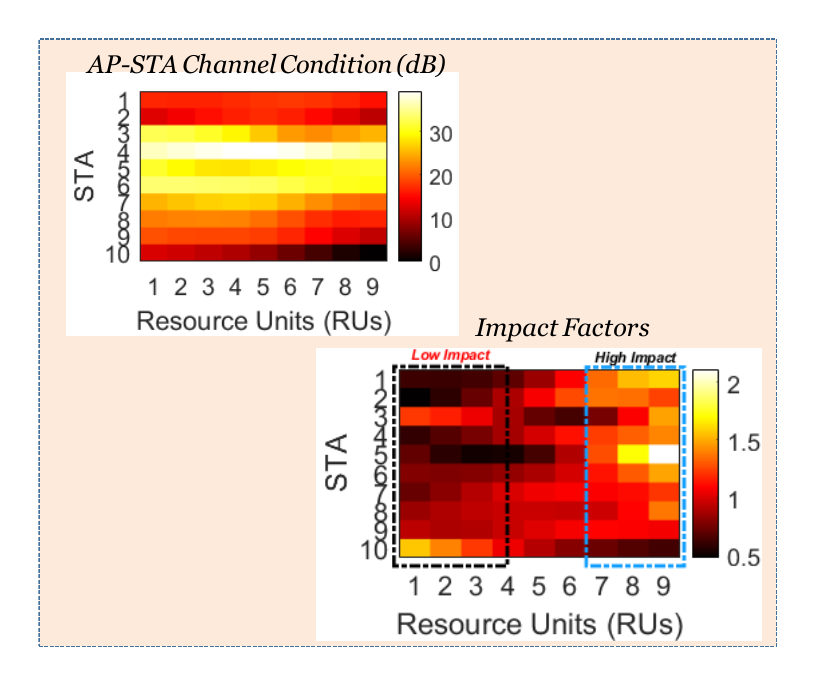 }%
\label{obs}}
\hfil
\caption{The (a) MU-MIMO-OFDMA-enabled WLAN where the channel between the AP and a STA is considered an agent that observes (b) the AP-STA channel conditions and the corresponding impact factors of all agents over the nine RUs.}
\label{env_obs}
\end{figure}

Consider the joint MU-MIMO-OFDMA-enabled WLAN, where an $M$-antenna WLAN AP is in charge of scheduling a set of $\mathcal{N}$ single-antenna STAs over a set of $\mathcal{R}$ RU configurations for $T$ time-slotted UL-SAs. Fig. \ref{wlan_env} illustrates the indoor WLAN environment. In this study, we consider the RU configuration with nine 26-tone RUs (i.e., $|\mathcal{R}|=R=9 $ is the granularity for resource assignment). In such a WLAN, AP initiates the UL-SA by transmitting a Trigger Frame (TF) to assign the RUs, Modulation and Coding Schemes (MCSs) to $|\mathcal{N}|= N$ STAs based on the channel conditions $\mathbf{h} \in \mathbb{C}^{1 \times M}$ between itself and each STA. Fig. \ref{obs} illustrates the $R$ AP-STA channels using heatmap plots, where brighter intensity indicates better channel conditions. Each STA distributes its transmit power budget $P_{total}$ over the assigned RUs to send data packets. Note that the RU assignments schedule groups of STAs to transmit concurrently and each group is referred to as the MIMO user group. Then, the transmitted symbol from the $i^{th}$ STA over the $l^{th}$ RU for the $t^{th}$ UL-SA, $x_{i,l,t}$, is received at the AP as: $\mathbf{y_{i,l,t}} = \mathbf{\hat{h}}_{i,l,t}x_{i,l,t} + \sum_{j\in\mathcal{N} \atop j\neq i}\mathbf{\hat{h}}_{j,l,t}x_{j,l,t} + \mathbf{n}_{l,t}, i \in \mathcal{N}, l \in \mathcal{R}$, where $\mathbf{n}_{l,t}$ $\sim \mathcal{C}\mathcal{N}(0, \sigma^2)$ is the noise at the receiver and $\mathbf{\hat{h}} $ is the MIMO channel estimated utilizing the orthogonal reference symbols that the AP assigned to each STA using the TF and organized as $\mathbf{\hat{H}}_{l,t} \in \mathbb{C}^{M \times N}$. Note that $\mathbf{\hat{h}}  = \sqrt{p_{i,l,t}}*\mathbf{h}$, where $p_{i,l,t}$ is the $i^{th}$ STA's distribution of $P_{total}$ over the $l^{th}$ RU for the $t^{th}$ UL-SA.

To decode the received signal, AP constructs a linear decoder, such as Zero Forcing (ZF) or Minimum Mean Squared Error (MMSE). We employ the MMSE-based decoder that performs successive interference cancellation (SIC) by spatially projecting each of the $N$ STAs' signals in an optimal direction away from the other $N-1$ interfering signals \cite{Knightly}. Denote $\mathbf{w}_{i,l,t}$ as the weight coefficient for the MMSE-SIC decoding filter, $\gamma_{i,l,t}$ as the Signal-to-Interference and Noise Ratio (SINR), and $C_{i,l,t}$ as the achievable data rate of the $i^{th}$ STA over the $l^{th}$ RU for the $t^{th}$ UL-SA. The weight coefficients are estimated using the expression \cite{userGrouping}: $\mathbf{w}_{i,l,t} = ((\mathbf{\hat{H}}_{l,t}\mathbf{\hat{H}}_{l,t}^H + \sigma^2\mathbf{I}_{M_{ap}})^{-1} \mathbf{\hat{h}}_{i,l,t}), \\ i \in \mathcal{N}, l \in \mathcal{R}$. The SINR $\gamma_{i,l,t}$ is defined as:  
\begin{equation}\label{SINR}
 \gamma_{i,l,t} = \frac{\lvert \mathbf{w}^{H}_{i,l,t} \mathbf{\hat{h}}_{i,l,t}\rvert^2}{ 
 \mathbf{w}^{H}_{i,l,t} (\mathbf{\hat{H}}_{j,l,t}\mathbf{\hat{H}}^{H}_{j,l,t} + \sigma^2\mathbf{I}_{M}) \mathbf{w}_{i,l,t}}, \\ i \in \mathcal{N}, l \in \mathcal{R}, 
\end{equation}
\noindent where $\mathbf{\hat{H}}_{j,l,t}=\mathbf{\hat{H}}_{l,t}$ with the $i^{th}$ column removed, and the terms in the numerator and denominator of \eqref{SINR} correspond to the signal and interference-plus-noise power \cite{userGrouping}, respectively. Then, for a given SINR, the $i^{th}$ STA's data rate can be estimated using MCS-based rate approximations. Denote $\hat{C}$ as the approximate MCS-based achievable rate, which is estimated using: ${\hat{C}(\gamma)} \leq {{m}_k}{\gamma} + {b_k},  {k \in \mathcal{K}},$ where ${m_k}$ is the gradient, $b_k$ is the constant, and $K \triangleq \{1,2,\cdots,13\}$ is the set of operating points for linear approximation. The values of ${m_k}$ and $b_k$ for approximating the MCS-based data rates are discussed in \cite{userGrouping}. As the approximated rate region is non-linear, we consider an upper bound on the SINR, denoted as $\gamma_{o}$, and the corresponding maximum achievable data rate ${S}$.

\subsubsection{Problem Formulation}
We define the SRA for UL-SA as an optimization problem with the objective to maximize the throughput (i.e., rate-sum of the UL-SA). Denote $B_l$ as the channel bandwidth of the $l^{th}$ RU, $\nu$ as the binary RU assignment variable, where  $\nu_{i,l,t} =1$ if the $l^{th}$ RU is assigned to the $i^{th}$ STA for the $t^{th}$ UL-SA, and $0$ otherwise. Then, we formulate the SRA problem for the UL-SA of the joint MU-MIMO-OFDMA-enabled WLAN as follows: 

\begin{equation}
	\label{sum_rate_3}
	 \max_{\tilde{\nu}}\sum_{i \in \mathcal{N}} C_{i,t} 
\end{equation}

s.t.  
\begin{equation}
    \label{constraint_4a}
    C_{i,l,t} \leq {B_l}{\nu_{i,l,t}} { ( {{m}_k} {\gamma_{i,l,t}} + {b_k} )}, \ {i \in \mathcal{N}}, {l \in \mathcal{R}}, {k \in \mathcal{K}}, 
\end{equation}

\begin{equation}
	\label{constraint_4b}
	  C_{i,t} \leq \sum_{l \in \mathcal{R}} C_{i,l,t} \ , \ {i \in \mathcal{N}} 
\end{equation}

\begin{equation}
	\label{constraint_4c}
	  \sum_{i \in \mathcal{N}} \nu_{i,l,t} \leq {M} \ , \ {l \in \mathcal{R}} 
\end{equation}

\begin{equation}
	\label{constraint_4d}
	  \sum_{l \in \mathcal{R}} {p_{i,l,t}} \leq {P_{total}} \ , \ {i \in \mathcal{N}} 
\end{equation}


\begin{equation}
	\label{constraint_4e}
    {0} \leq C_{i,l,t} \leq \nu_{i,l,t}{S}. 
\end{equation}

In \eqref{sum_rate_3}, $\tilde{\nu}$ is the optimization variable, while the transmit power budget $P_{total}$ of the STAs, and the $M$ antennas on the AP are given. Constraints \eqref{constraint_4a} and \eqref{constraint_4b} defines the $i^{th}$ STA's achievable data rate for the $t^{th}$ UL-SA. Constraint \eqref{constraint_4c} is the MIMO user grouping/spatial constraint that limits the number of STAs that can simultaneously occupy the same RU. Let $g_{l,t}$ denote the size of the MU-MIMO user group over the $l^{th}$ RU for the $t^{th}$ UL-SA. Then, constraint \eqref{constraint_4c} is met if $g_{l,t} \leq M$. Constraint \eqref{constraint_4d} defines the distributed power allocation by the STAs for the $t^{th}$ UL-SA. Lastly, Constraint \eqref{constraint_4e} limits the optimization search space. 

To attain the objective in \eqref{sum_rate_3}, we assume the WLAN AP utilizes prior knowledge of the channel conditions (i.e., $\mathbf{h}$) to schedule spatially compatible groups of STAs with uncorrelated channel conditions over the available RUs. However, for the UL-SA, STAs distribute their transmit power budget among the assigned RUs. This power allocation determines the UL channel conditions (i.e., $\mathbf{\hat{h}}$), SINR, and the UL rate-sum. Therefore, the AP cannot evaluate the impact of its SRA strategy prior to the UL-SA, and must perform a computationally expensive exhaustive search over all possible combinations of MIMO user groups to effectively assign the RUs.

\section{Scheduling and Resource Allocation using Multi-Agent Language Model-Based System}\label{maxlm}

To simplify the optimization of the SRA problem for the UL-SA of a joint MU-MIMO-OFDMA-enabled WLAN, we propose a Multi-Agent small/medium-sized Language Model (MAxLM)-based system. To facilitate the proposed MAxLM-optimized SRA, we introduce the AI-assisted \textbf{Wi}reless \textbf{S}ystems \textbf{E}ngineering and \textbf{R}esearch (WiSER) platform. Fig. \ref{workflow_modules}(a) illustrates the WiSER modules and the workflow of the MAxLM-optimized SRA.

\subsection{Framework of the Proposed MAxLM-optimized SRA}

The proposed MAxLM-optimized SRA adopts the multi-agent framework to solve \eqref{sum_rate_3} by modeling the varying states of the AP-STA connections in the WLAN environment as a Markov Decision Process (MDP) problem and defining the MDP components: the agent, state space, action space, and environmental feedback.

\subsubsection{Agent} We consider the wireless channel between the AP and a STA as an agent, referred to as the AP-STA UL connection. Fig. \ref{wlan_env} illustrates the agents in the indoor WLAN environment. Each agent is a candidate for multi-channel UL access and can independently be assigned the $R$ RUs in any one of the $\sum^{R}_{l=0} {R\choose l}$ ways. In contrast, as a single agent, the AP has to choose from $\sum^{M}_{m=1} {N\choose m}$ possible combinations of MIMO user groups to schedule $M$ spatially compatible STAs over the $R$ RUs. Therefore, the multi-agent realization simplifies the optimization of the SRA problem by reducing the agent's action space.
\subsubsection{State Space} An agent's state space $\mathcal{S}$ consists of agent's observations of the environment. Therefore, we include channel gains of the $N$ agents, denoted as $\boldsymbol{\zeta}$ and organized as: $\boldsymbol{\zeta}_{l} = [ ||\mathbf{h}_{1,l}||_F, ||\mathbf{h}_{2,l}||_F, \cdots, ||\mathbf{h}_{N,l}||_F]^T$, where $\mathbf{h} \in \mathbb{C}^{1 \times M}$ is the $i^{th}$ STA's channel state over the $l^{th}$ RU and $||\cdot||_F$ denotes the Frobenius norm. Recall that the UL throughput is affected by spatial interference among the STAs in MIMO user groups scheduled for the UL-SA. So, we define and include in $\mathcal{S}$ the agent's impact factor $\eta$, which is a measure of an agent's spatial compatibility with the other agents. Let $\boldsymbol{\eta}_{t} \in \mathbb{R}^{N \times R}$ denote the matrix that consists of the impact factors for the $t^{th}$ UL-SA. Then, $\eta_{i,l,t} = R\left(\frac{ \sum^{N-1}_{k=1} \tilde{\zeta}_{k,l,t} }{ \sum^{R}_{l=1}\sum^{N-1}_{k} \tilde{\zeta}_{k,l,t}} \right) $, where $\tilde{\zeta}_{k,l,t}$ is obtained from the normalized channel gain matrix  $\boldsymbol{\tilde{\zeta}}_{i,t}= \frac{\boldsymbol{\zeta}_{j,t}}{\boldsymbol{\zeta}_t}$ (i.e., $\ i \neq j, j = \{1,2,\cdots,N-1\}$ and $ \boldsymbol{\zeta}_t = [ \boldsymbol{\zeta}_{1}, \boldsymbol{\zeta}_{2}, \cdots, \boldsymbol{\zeta}_{R}] $) \cite{userGrouping}. Fig. \ref{obs} illustrates the agents' impact factors for the given AP-STA channel conditions. It can be seen in Fig. \ref{obs} that agents are assigned higher impact factors over RUs $7$ to $9$ as their channel conditions over those RUs are somewhat uncorrelated. Additionally, result of the $i^{th}$ agent's action in the past state is included in the state space, represented using a set of feedback $\mathcal{F}$, denoted as $\mathbf{f}_{t-1} \in \mathcal{F}$. Therefore, an agent's observations for the state $s_{i,t} \in \mathcal{S}$ are expressed as:
\begin{equation}
    s_{i,t} = \{ \boldsymbol{\zeta}_{t}, \boldsymbol{\eta}_{t}, \mathbf{f}_{t-1} \}
\end{equation}

\subsubsection{Action Space} The action space consists of the agent's choices, i.e., the possible combinations of assigning the $R=9$ RUs. Let $\mathcal{K}=\{0,1,\cdots,R\}$ denote assigning $k \in \mathcal{K}$ of the $R$ RUs. Let $\mathcal{U}_{k}$ and $\mathbf{a}_{i,t} \in \{0,1\}^{1 \times R}, i \in \mathcal{N}$ denote the set containing all possible RU assignment strategies resulting from assigning $k$ of the $R$ RUs and $i^{th}$ agent's RU assignments for the $t^{th}$ UL-SA, respectively. Then, to assign all of the nine RUs, $\mathbf{a}_{i,t}=[\nu_{i,1,t},\nu_{i,2,t}, \cdots,\nu_{i,R,t}] \in \mathcal{U}_{9}$, where $\sum_{l \in \mathcal{R}} \nu_{i,l,t}= 9$ and $|\mathcal{U}_k|=$ $R \choose k$. Therefore, the agent's action space is defined to include all the possible RU assignment strategies in the set $\mathcal{A}$, expressed as:    
\begin{equation}\label{action_space}
\mathcal{A} = \{ \mathcal{U}_{k}\ | \ k \in \mathcal{K} \}.
\end{equation}

\subsubsection{Feedback} To avoid aggregating redundant context and minimize inference latency, we limit agents' feedback to system notification indicating success or error in parsing the xLM's intent. Then, let $\mathcal{F} = \{ F_{1}, F_{2}, \cdots, F_{J} \}$ consist of the $J$ \textit{string}-valued instances that characterize the parsing statuses and represent the agents' feedback. Note that we do not include the past actions and data rates as part of the feedback because the xLM's SRA intent for the past may not apply for the current state of a dynamically changing WLAN. Moreover, the ACM procedure (detailed in Section \ref{maxlm_wiser}) enables the xLM to effectively allocate RUs using $\boldsymbol{\zeta}_t$ and $\boldsymbol{\eta}_t$.

\begin{figure}[t]
\centering
	{\includegraphics[width=1\linewidth]{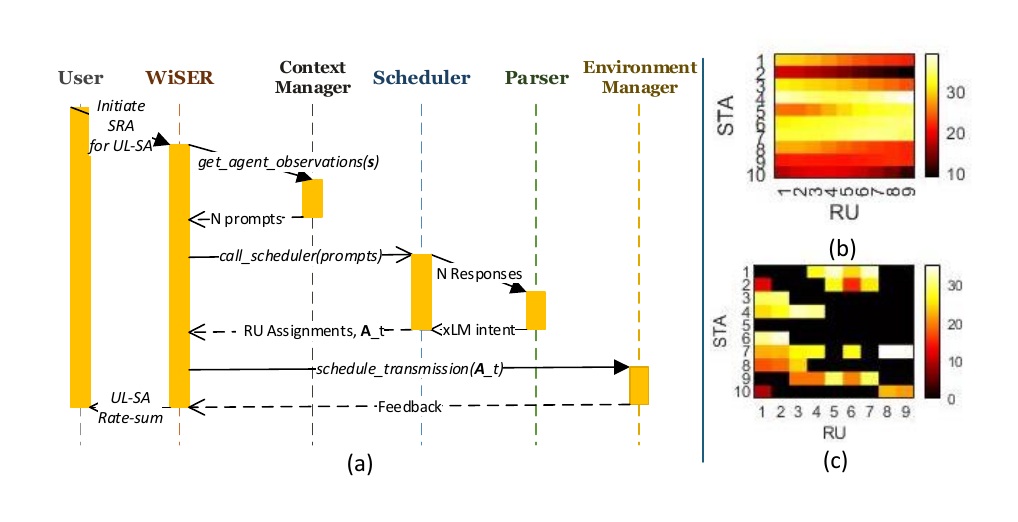}}
    \caption{The MAxLM-optimized SRA: (a) workflow illustrated using a UML sequence diagram. An example case where an 8-antenna AP schedules 10 stations, for the WLAN state with the (b) AP-STA channel conditions shown over the nine RUs, using (c) the RU assignments inferred by Mistral-NeMo:12b.}
    \label{workflow_modules}
\end{figure}

\subsection{MAxLM-optimized SRA Algorithm and WiSER Modules}\label{maxlm_wiser}

When initialized, WiSER runs the MAxLM-optimized SRA algorithm in sequential steps. First, each agent shares its observations of the WLAN environment's state $s_{i,t} \in \mathcal{S}$ with WiSER's \textit{Context Manager} module, which generates $N$ prompts for the $t^{th}$ UL-SA. Second, these prompts are loaded asynchronously onto an xLM's context window using WiSER's \textit{Scheduler} module that enables the xLM to select the action $a \in \mathcal{A}$ for each agent in a decentralized fashion. Third, WiSER's \textit{Parser} module parses the xLM's SRA intent, converts the assignments into binary representation and stores them in the action matrix, denoted as $\mathbf{A}_t$. Note that the decentralized SRA poses the challenge of potentially scheduling more than $M$ agents over the same RU. Therefore, a \textit{self-correction} step revokes the RU assignments of all the agents occupying any RU in violation of constraint \eqref{constraint_4c}, and issues the revised assignments $\mathbf{A}^{'}_t$. Fourth, WiSER's \textit{Environment Manager} module enables the AP to initiate the $t^{th}$ UL-SA and process the received signal to estimate the SINR and data rates of each STA using \eqref{SINR} and \eqref{constraint_4b}, respectively. 

As illustrated using the Unified Modeling Language (UML) sequence diagram in Fig. \ref{workflow_modules}(a), the workflow of MAxLM-optimized SRA is implemented as a graph \cite{langgraph} and the methods \textit{get\_agent\_observations($s$), \textit{call\_scheduler($prompts$)}}, and \textit{schedule\_transmission($A_t$)} are the graph nodes. The flow of information between these nodes is supported using a shared state schema. Fig. \ref{workflow_modules}(b) and Fig. \ref{workflow_modules}(c) show the AP-STA channel conditions for a given state $s \in \mathcal{S}$ and the corresponding RU assignments inferred by the xLM (Mistral-NeMo:12b) using heatmap plots. The bright cell intensity in the heatmap plots represents good channel conditions and assignment of an RU. In what follows, we discuss the WiSER modules and summarize the MAxLM-optimized SRA algorithm in Algorithm \ref{algo1}.

\subsubsection{Context Manager} The \textit{Context Manager} module performs ACM to prompt the xLM with an accurate representation of the changing context behind the varying states of the dynamic WLAN environment. Fig. \ref{wiser_prompt} illustrates the modular prompts that define the xLM's role and task: to follow either an intent-based SRA strategy, such as the best channel quality (BCQ)-based SRA, or the MAxLM-optimized SRA strategy. We recognize the most effective MAxLM-optimized SRA strategy as the one that prompts the xLM to infer an agent's RU assignments by intelligently prioritizing among the agent's attributes: channel strength, spatial compatibility, and how each agent's attributes compare with those of the other agents. 

Therefore, the $i^{th}$ agent's prompt is updated with the following semantic analyses of the agent's attributes: \textit{i) Agent\_$i$ is/is not one of the $M$ strongest agents; ii) Agent\_$i$ is/is not one of the $M$ most compatible agents; iii) There are $x$ agents that are stronger and more compatible with others than Agent\_$i$.} Fig. \ref{wiser_prompt} shows the prompt Templates 1 and 2, which provide semantic and numerical representations of the agent's observations, respectively. For Template 1, these analyses of the agent's attributes over the $R$ RUs are stored as \textit{string}-valued instances, utilizing condition-based programming, in the following arrays: \textit{agent\_strength}, \textit{agent\_compatibility}, and  \textit{agent\_comparison} \cite{wiser_github}. These arrays are then injected into the prompt. Thus, ACM enables the xLM to anticipate an effective allocation of RUs since the AP is unable to assess the impact of the selected SRA strategy prior to the UL-SA.
\begin{algorithm}[!t]
 \caption{\strut MAxLM-optimized SRA}
 \begin{algorithmic}[1]
 \label{algo1}
 \renewcommand{\algorithmicrequire}{\textbf{Input:}}
 \renewcommand{\algorithmicensure}{\textbf{Output:}}
 \REQUIRE Agents' observations of the WLAN environment
 \ENSURE  Resource allocation for the UL-SA 
 \STATE \textit{initialize}: $T, N, R, M, $ initial state $s_{i,t} \in \mathcal{S}$, $t=0$
   
   \FOR {$t = 1$ to $T$} 
        \FOR {$i = 1$ to $N$} 
            \STATE share agent's observations 
            \STATE generate $i^{th}$ agent's prompt  
        \ENDFOR 
        \STATE select action $a \in \mathcal{A}$ asynchronously
        \STATE parse RU assignments and trigger $t^{th}$ UL-SA
        \STATE estimate SINR by \eqref{SINR}, data rate  by \eqref{constraint_4b}
    \STATE observe the next state, $s_{i,t+1}$
    \STATE update $s_{i,t} \xleftarrow{} s_{i,t+1}$
    \ENDFOR 
\end{algorithmic} 
\end{algorithm}

\subsubsection{Scheduler and Parser} 
\begin{figure}[!t]
\centering
	{\includegraphics[width=1\linewidth]{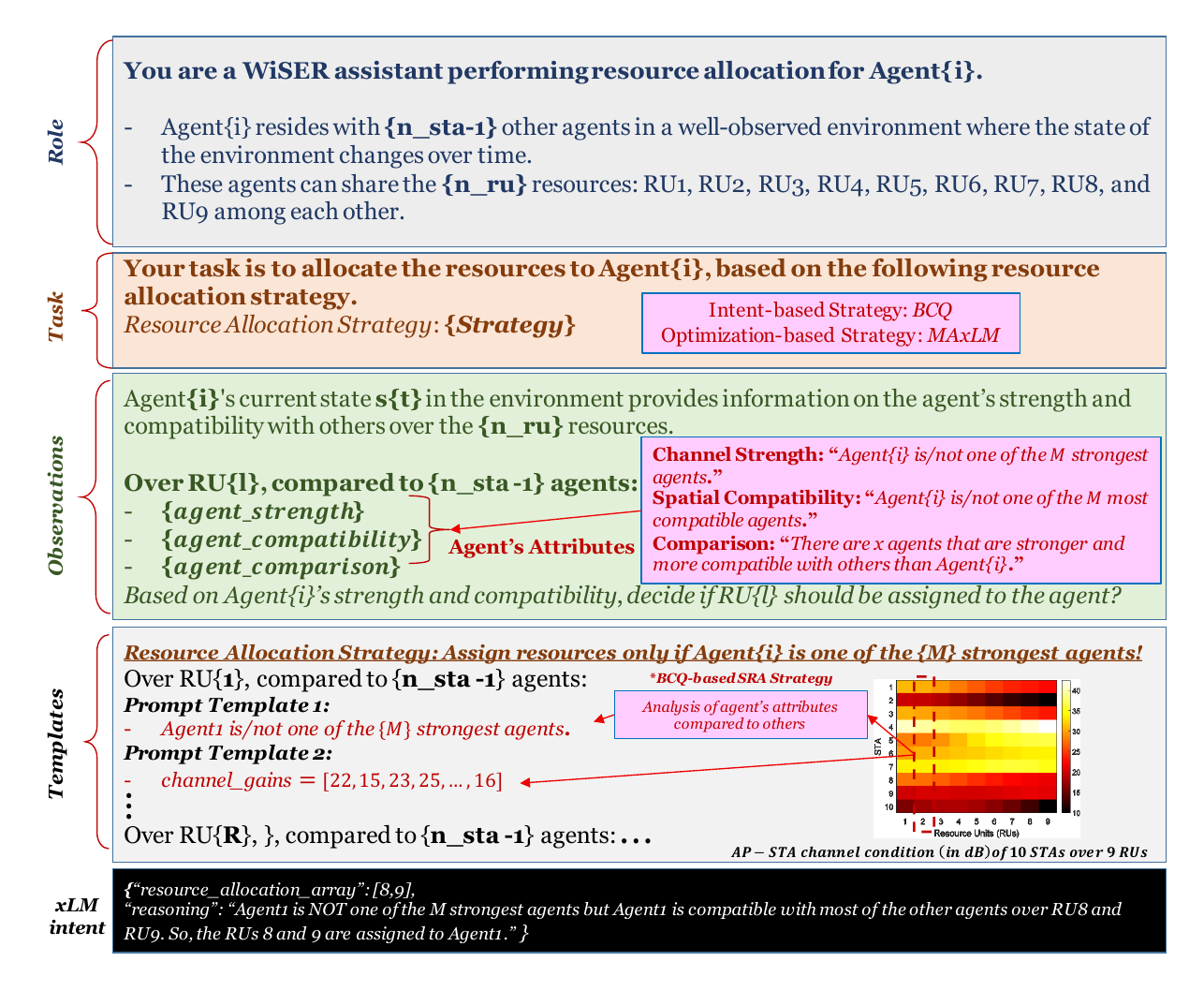}}
		\caption{Modular structure of the prompts showing the role, task, analysis of agents' observations for two templates, and an example of the xLM's response.}
		\label{wiser_prompt}
\end{figure}

The \textit{Scheduler} module streamlines the SRA process for the UL-SA by loading the $N$ prompts onto the context window of the xLM asynchronously. As shown in Fig. \ref{wiser_prompt}, the xLM is instructed to express its SRA intent for each agent in the JavaScript Object Notation (JSON) format. The JSON-formatted output contains an agent's RU assignments and the reasoning behind the xLM's intent. Subsequently, WiSER's \textit{Parser} module parses the xLM's SRA intent as the binary RU assignments for the UL-SA. Currently, WiSER supports the instruct-based model Llama3.1:8b, the reasoning-based models Mistral-NeMo:12b and Gemma:12b. The xLMs are obtained from \cite{Ollama_models} and hosted on a server equipped with the NVIDIA GeForce RTX 2080 Ti GPU (12GB VRAM). We observe that the instruct- and reasoning-based models require approximately $30-45$ and $45-60$ seconds to perform SRA among $10$ stations, respectively. For practical deployment, we assume the AP executes the MAxLM-optimized SRA by connecting to the external server.

\subsubsection{Environment and File Manager} The \textit{Environment Manager} is responsible for simulating or providing access to the agent's environment, as well as hosting the \textit{self-correction} step to validate RU assignments. Currently, the module utilizes the MATLAB engine \cite{matlab_engine} to simulate the UL-SA of the MU-MIMO-OFDMA-enabled WLAN and is designed to enable convenient integration of WiSER's modules with external research tools/platforms. Lastly, the \textit{File Manager} module allows WiSER to store and load SRA strategies, WLAN channel data, and xLM responses.

\section{Performance Analysis}\label{results}

We design experiments to evaluate the performance of MAxLM-based SRA for the UL-SA of the joint MU-MIMO-OFDMA-enabled WLAN. We consider the BCQ-based scheme as the benchmark technique, which assigns the $R=9$ RUs to $M$ WLAN single-antenna STAs with the best channel gains/strengths. The UL channel conditions between the AP and the STAs are modeled using the IEEE TGax Indoor channel model \cite{Matlab_channel} with Model-B delay profile, and are further discussed in \cite{userGrouping}. We consider the STAs to be within 15 m of the WLAN AP but their locations vary randomly across UL-SAs. We collect channel data for the cases where a $4$-antenna and $8$-antenna WLAN AP schedules the STAs in each of the scenarios with $10$ and $20$ stations in the WLAN. For each of the cases and scenarios, the channel data is gathered for $1200$ test episodes, where each episode consists of $T=50$ UL-SAs.

\subsection{Analysis of Prompt Templates}

We tasked the xLMs to assist the $4$-antenna AP in scheduling $10$ stations, using the BCQ-based SRA strategy, for the $50$ UL-SAs in two randomly selected test episodes. We use the prompt template 1 (PT1) that provides the agent with semantic analysis of the agent's attributes versus prompt template 2 (PT2), which prompts the xLM with $1 \times N$ dimensional array containing the data on UL channel conditions over each of the $R$ RUs (see Fig. \ref{wiser_prompt}). For the case with the $4$-antenna AP, the BCQ-based SRA assigns $M=N=4$ STAs with the best channel quality over the nine RUs. These assignments are considered as the actual/true assignments. 

\begin{figure}[!t]
\captionsetup[subfloat]{labelformat=simple}
\centering
\subfloat[]{\includegraphics[width=0.5\columnwidth]{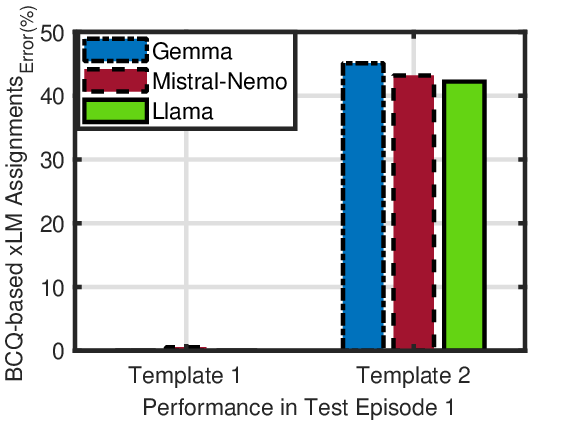 }%
\label{pt_errors_bar1}}
\hfil
\subfloat[]{\includegraphics[width=0.5\columnwidth]{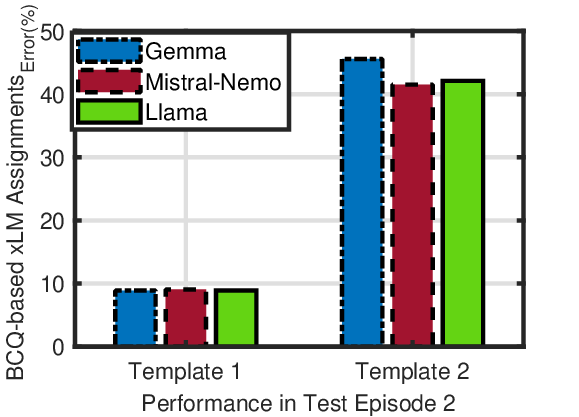}%
\label{pt_errors_bar2}}
\hfil

\caption{Assignment errors made by the xLMs performing the BCQ-based SRA for the $4$-antenna AP scheduling 10 stations in (a) Test Episode 1, and (b) Test Episode 2 using the prompt Templates 1 and 2.}
\label{pt_errors}
\end{figure}

Fig. \ref{pt_errors_bar1} and Fig. \ref{pt_errors_bar2} show the performance of the xLMs for the two test episodes in terms of errors, represented by the false positives (FP) and false negatives (FN) (i.e., $\nu^{actual}_{i,l,t} = 1$ but $\nu^{inferred}_{i,l,t} = 0$ and $\nu^{actual}_{i,l,t} = 0$ but $\nu^{inferred}_{i,l,t} = 1$). Precisely, the errors are estimated using: $Error_{MAxLM-BCQ}= \frac{(FP + FN)}{(R*N*T)}$. Fig. \ref{pt_errors_bar1} shows that when the xLMs are prompted with PT1, Mistral-NeMo was the only xLM that incurred errors, with an error rate of $0.5\%$. In contrast, with PT2, all the xLMs yield an average error of $43\%$. From Fig. \ref{pt_errors_bar2}, the three xLMs yield a low average error of $8.5\%$ with PT1 compared to high average error of $38\%$ with PT2. Therefore, we use PT1 to conduct the rest of the experiments.

\subsection{Analysis of UL-SA Rate-sum with MAxLM-optimized SRA}

\begin{figure}[!t]
\captionsetup[subfloat]{labelformat=simple}
\centering
\subfloat[]{\includegraphics[width=0.45\columnwidth]{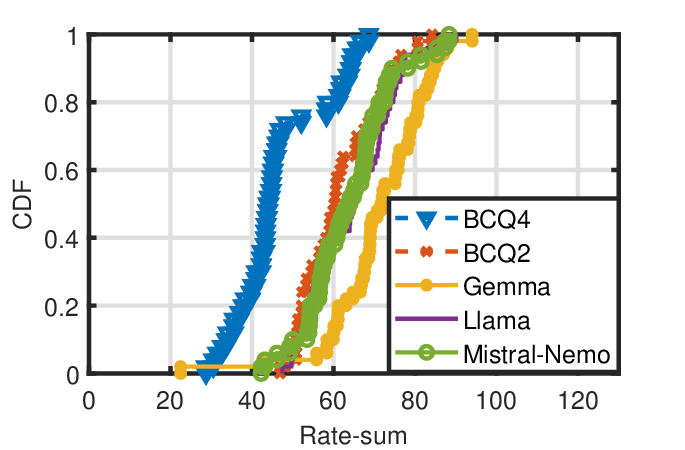}%
\label{cdf10usr_4ant}}
\hfil
\subfloat[]{\includegraphics[width=0.45\columnwidth]{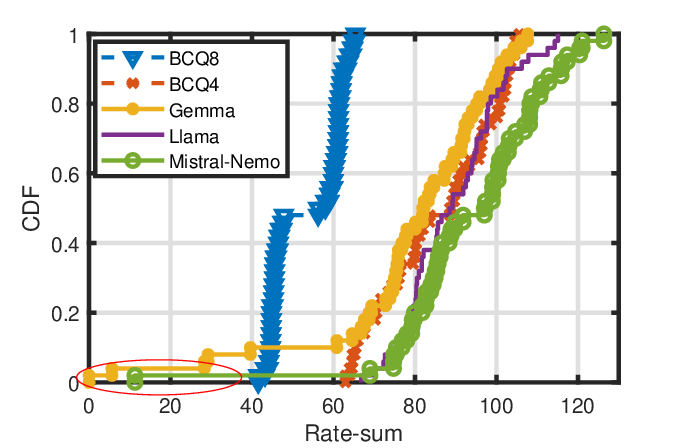}%
\label{cdf10usr_8ant}}
\hfil
\subfloat[]{\includegraphics[width=0.45\columnwidth]{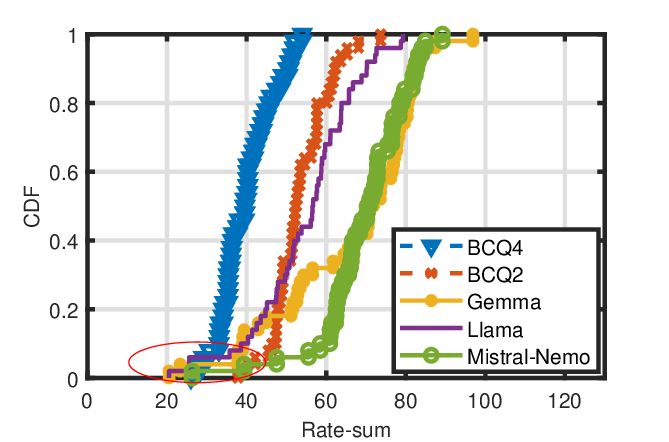}%
\label{cdf20usr_4ant}}
\hfil
\subfloat[]{\includegraphics[width=0.45\columnwidth]{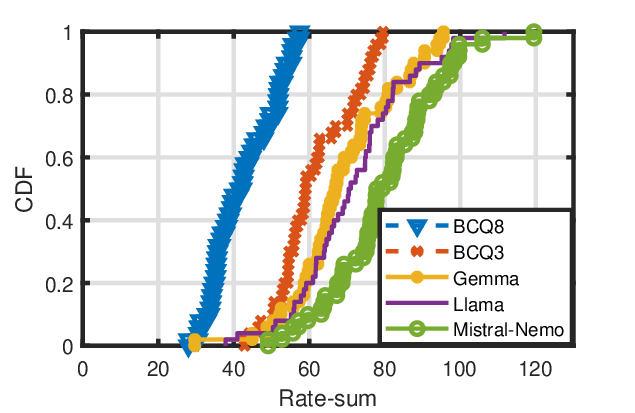}%
\label{cdf20usr_8ant}}
\hfil

\caption{CDF of the UL-SA rate-sum utilizing the MAxLM-optimized SRA for: (a) 10 stations with a 4-antenna AP, (b) 10 stations with an 8-antenna AP, (c) 20 stations with a 4-antenna AP, and (d) 20 stations with an 8-antenna AP.}
\label{cdf_rates}
\end{figure}

We compare the UL rate-sum achieved by the MAxLM-optimized SRA with that of the BCQ-based SRA, for the $50$ UL-SAs in a randomly selected test episode. As the evaluation metric, we select the average performance gain, which is expressed as: $Gain = \frac{\sum^{50}_{t=1} \sum^{N}_{i=1} (C^{xLM}_{i,t} - C^{BCQ}_{i,t}) }{\sum^{50}_{t=1} \sum^{N}_{i=1} (C^{BCQ}_{i,t})} \times 100$, where $C^{xLM}$ and $C^{BCQ}$ represent the UL rate-sum achieved by the MAxLM-optimized and the BCQ-based SRA, respectively. Fig. \ref{cdf10usr_4ant} and Fig. \ref{cdf10usr_8ant} illustrate the cumulative distribution function (CDF) of the UL-SA rate-sum achieved by the $4$- and $8$-antenna AP scheduling $10$ stations with the MAxLM-optimized RU assignments.

It can be seen in Fig. \ref{cdf10usr_4ant} that Gemma outperforms the other xLMs by achieving a $16\%$ gain in performance over BCQ2, which assigns $N=2$ STAs with the best channel quality over the nine RUs. Both Llama and Mistral-NeMo attain similar gains of $5\%$ and $3\%$ over BCQ2, respectively. In contrast, Fig. \ref{cdf10usr_8ant} demonstrates that Mistral-NeMo and Llama achieve performance gains of $10\%$ and $4\%$ over BCQ4, which assigns $N=4$ STAs with the best channel quality over the nine RUs. Whereas, Gemma incurs a performance loss of $8\%$ as it yields low UL rate-sums for a few of the $50$ UL-SAs. It is observed that Gemma-optimized SRA violated the MIMO spatial constraint \eqref{constraint_4c} for the UL-SAs marked by the red ellipse in Fig. \ref{cdf10usr_8ant}. As a result the MAxLM's self-correction step revoked RU assignments for all RUs violating spatial constraint \eqref{constraint_4c}, which decreased the UL rate-sum.  

Fig. \ref{cdf20usr_4ant} and Fig. \ref{cdf20usr_8ant} show the CDF of the UL rate-sum achieved by the $4$- and $8$-antenna AP scheduling $20$ stations with the MAxLM-optimized RU assignments. It is seen from Fig. \ref{cdf20usr_4ant} that Mistral-NeMo outperforms the other xLMs by achieving a performance gain of $30\%$ over BCQ2. Gemma closely follows with a gain of $23\%$, and Llama achieves a gain of $2.5\%$. Llama's low performance gain is due to its adopting a conservative resource allocation strategy similar to that of BCQ2. From Fig. \ref{cdf20usr_8ant}, it is evident that Mistral-NeMo outperforms both the benchmark and the other xLMs by maintaining a performance gain of $30\%$ over BCQ3. Overall, from Fig. \ref{cdf_rates}, it can be seen that Mistral-NeMo performs consistently across all the cases and WLAN scenarios. 

\subsection{Analysis of MIMO User Group Size Distribution}
\begin{figure}[!t]
\captionsetup[subfloat]{labelformat=simple}
\centering
\subfloat[]{\includegraphics[width=0.45\columnwidth]{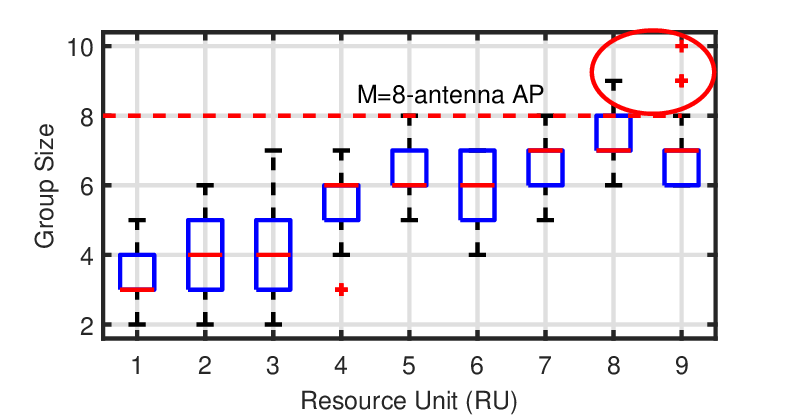}%
\label{grpS_10usr_8ant}}
\hfil
\subfloat[]{\includegraphics[width=0.45\columnwidth]{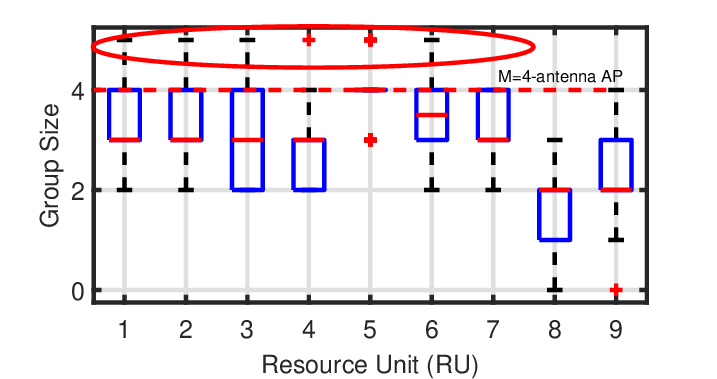}%
\label{grpS_20usr_4ant}}
\hfil
\caption{Analysis of Gemma-optimized SRA's MIMO user group size distribution for the scenarios and cases where (a) 10 stations are scheduled by the 8-antenna AP, and (b) 20 stations are scheduled by the 4-antenna AP.}
\label{grpSizes}
\end{figure}

It is seen in Fig. \ref{cdf10usr_8ant} and Fig. \ref{cdf20usr_4ant} that the Gemma-optimized SRA yields notably poor UL rate-sum for some of the $50$ UL-SAs in the test episode. Therefore, we analyze the impact of Gemma's SRA strategy by examining the MIMO user group size distributions over the $R=9$ RUs. Fig. \ref{grpS_10usr_8ant} illustrates the group size distributions for the WLAN scenario and case where $10$ stations are scheduled with the $8$-antenna AP. In this case, Gemma-optimized SRA assigns $6$ STAs, on average, over the nine RUs. However, over the RU8 and RU9, Gemma assigns $N > M$ STAs violating the MIMO user grouping constraint \eqref{constraint_4c} (as seen within the red ellipse), which prompts revising the RU assignments by prohibiting most of the $N=10$ STAs from participating in the UL-SA. Thus, one of the $50$ UL-SAs in the test episode yields a zero rate-sum. 

Fig. \ref{grpS_20usr_4ant} shows the group size distributions for the $4$-antenna AP scheduling $20$ stations. From Fig. \ref{grpS_20usr_4ant}, it can be seen that the Gemma-based SRA strategy fails to satisfy the MIMO user grouping constraint \eqref{constraint_4c} over the RUs $1-6$, for $16$ of the $50$ UL-SAs. As a result, the revised RU assignments drop several STAs scheduled for the UL-SA, which lowers the UL rate-sum. Note that both Gemma and Mistral-NeMo achieved a performance gain over than the benchmark, BCQ2. With BCQ-based SRA, only a fixed set of STAs are scheduled over the nine RUs. In contrast, the MAxLM-optimized SRA assigns a group of spatially compatible STAs over a varying number of the RUs to maximize the UL-SA rate-sum.

\section{Conclusions}\label{conc}
In this paper, we have addressed the SRA problem for the UL-SA of a joint MU-MIMO-OFDMA-enabled WLAN, and proposed the MAxLM-optimized SRA that leverages the advances in generative AI. To execute MAxLM-optimized SRA, WiSER was introduced as a robust platform for autonomous WLAN resource management, and the functionality of its modules was discussed. Numerical results confirmed that the MAxLM-optimized SRA using Mistral-NeMo:12b as the xLM can achieve a $30\%$ performance gain in terms of UL throughput over the benchmark techniques. 


\vspace{12pt}


\begin{thebibliography}{00}
\bibitem{userGrouping45}R. Zhang, K. Xiong, Y. Lu, B. Gao, P. Fan, and K. B. Letaief, ‘‘Joint coordinated beamforming and power splitting ratio optimization in mu-miso swipt-enabled hetnets: A multi-agent ddqn-based approach,’’ IEEE Journal on Selected Areas in Communications, vol. 40, no. 2, pp. 677–693,
2021.

\bibitem{Raudl}A. Quadri and H. Li, ‘‘Resource allocation using deep learning in uplink 802.11 ax networks,’’ in GLOBECOM 2023-2023 IEEE Global Communications Conference. IEEE, 2023, pp. 4020–4025.
\bibitem{userGrouping33}R. Balakrishnan, K. Sankhe, V. S. Somayazulu, R. Vannithamby, and J. Sydir, ‘‘Deep reinforcement learning based traffic-and channel-aware ofdma resource allocation,’’ in 2019 IEEE Global Communications Conference (GLOBECOM). IEEE, 2019, pp. 1–6.

\bibitem{userGrouping47}Y. S. Nasir and D. Guo, ‘‘Multi-agent deep reinforcement learning for dynamic power allocation in wireless networks,’’ IEEE Journal on Selected Areas in Communications, vol. 37, no. 10, pp. 2239–2250, 2019.

\bibitem{userGrouping46}N. Naderializadeh, J. J. Sydir, M. Simsek, and H. Nikopour, ‘‘Resource management in wireless networks via multi-agent deep reinforcement learning,’’ IEEE Transactions on Wireless Communications, vol. 20, no. 6, pp. 3507–3523, 2021.

\bibitem{MAESTRO}I. Chatzistefanidis, A. Leone, and N. Nikaein, ‘‘Maestro: Llm-driven collaborative automation of intent-based 6g networks,’’ IEEE Networking Letters, vol. 6, no. 4, pp. 227–231, 2024.
\bibitem{LiINK}S. Jiang, B. Lin, Y. Wu, and Y. Gao, ‘‘Links: Large language model integrated management for 6g empowered digital twin networks,’’ in 2024 IEEE 100th Vehicular Technology Conference (VTC2024-Fall). IEEE, 2024, pp. 1–6.

\bibitem{LLMxapp}X. Wu, J. Farooq, Y. Wang, and J. Chen, ‘‘Llm-xapp: A large language model empowered radio resource management xapp for 5g o-ran,’’ 2025.

\bibitem{wiser_github}WiSER, “quadri-a/wiser,” GitHub. Accessed: Apr. 14, 2026. [Online]. Available: https://github.com/quadri-a/wiser

\bibitem{Knightly}O. Bejarano, S. Quadri, O. Gurewitz, and E. W. Knightly, ‘‘Scaling multi-user mimo wlans: The case for concurrent uplink control messages,’’ in 2015 12th Annual IEEE International Conference on Sensing, Communication, and Networking (SECON). IEEE, 2015, pp. 238–246.

\bibitem{userGrouping}A. Quadri, Q. Li, and H. Li, ‘‘User grouping and resource allocation for uplink of mu-mimo-ofdma-enabled wlan using multi-agent reinforcement learning,’’ IEEE Access, vol. 13, pp. 191 990–192 005, 2025.

\bibitem{langgraph}“Graph API overview.”. Accessed: Apr. 13, 2026. [Online]. Available: https://docs.langchain.com/oss/python/langgraph/graph-api

\bibitem{Ollama_models}“Ollama library.” Accessed: Apr. 13, 2026. [Online]. Available: https://ollama.com/library

\bibitem{matlab_engine}"Using MATLAB with Python." Accessed Aug. 1, 2025. [Online]. Available: https://www.mathworks.com/products/matlab/matlab-and-python.html
\bibitem{Matlab_channel}"Filter signal through 802.11ax multipath fading channel." Accessed Aug. 1, 2025. [Online]. Available: https://www.mathworks.com/help/wlan/ref/wlantgaxchannel-system-object.html

\end{thebibliography}
\end{document}